\documentclass[11pt,
]{article}
\usepackage[]{graphicx}
\usepackage{makeidx}
\newcommand{\mathsym}[1]{{}}

\newcommand{\bi}{\bibitem}
\newcommand{\be}{\begin{eqnarray}}
\newcommand{\ee}{\end{eqnarray}}
\newcommand{\rar}{\rightarrow}
\newcommand{\lrar}{\leftrightarrow}
\usepackage[]{caption}
\def\Journal#1#2#3#4{{#1} {\bf #2}, #3 (#4)}

\def\PISM{\em Pis'ma Z. Eksp. Teor. Fiz.}

\begin{document}

\begin{center}
{\bf \large{
CPT violation and particle-antiparticle asymmetry in cosmology 
}} \\ \vspace{0.5cm}
{\it A. D. Dolgov }
\\ \vspace{0.3cm}
Dipartimento di Fisica, Universit\`a degli Studi di Ferrara, 
I-44100 Ferrara, Italy \\
Istituto Nazionale di Fisica Nucleare, Sezione di Ferrara, 
I-44100 Ferrara, Italy \\
Institute of Theoretical and Experimental Physics, 113259 Moscow, Russia 

\begin{abstract}
General features of generation of the cosmological charge asymmetry in 
CPT non-invariant world are discussed. If the effects of CPT violation 
manifest themselves only in mass differences of particles and antiparticles, 
the baryon asymmetry of the universe hardly can be explained solely by 
breaking of CPT invariance. However, CPT non-invariant theories may lead to 
a new effect of distorting the usual equilibrium distributions. If this 
takes place, CPT violation may explain the baryon asymmetry of the universe.
\end{abstract}
\end{center}


The generally accepted mechanism of generation of cosmological charge 
asymmetry is based on three general principles put forward by 
Sakharov in 1967~\cite{sakharov}:\\
1. Nonconservation of baryon number.\\
2. Breaking of C and CP.\\
3. Deviation from thermal equilibrium. \\
It is established long ago by experiment that P, C, and CP are broken. Big bang
cosmology unambiguously states that massive particles should be out of thermal
equilibrium in the cosmological plasma. Nonconservation of baryons is predicted
by electroweak and grand unified theories and ``experimentally'' proven
by existence of our universe. So the Sakharov baryogenesis seems to be in a pretty
good shape, though some efforts are needed to obtain sufficiently large 
cosmological baryon asymmetry. In this connection it may be interesting to
explore other possibilities. For more details about the standard baryogenesis
and the list of references see e.g. reviews~\cite{ad-bs-rev}. 

Since all three symmetries C, P, and T are known to be broken, it is tempting
to explore consequences of breaking of the combined CPT symmetry. There is of
course a drastic difference between anyone of the single symmetry transformations 
or any pair of them and the combined action of the three. According to the 
celebrated CPT theorem~\cite{cpt}, any local Lorenz invariant theory with 
hermitian Hamiltonian, with positive 
definite energy or, better to say, with the canonical relation between spin and
statistics is invariant with respect to CPT transformation.
On the other hand, there are absolutely no theoretical arguments in favor of
invariance with respect to separate P, C, and T transformations and they are
indeed only approximate. If we trust CPT theorem then we should conclude
that any pair CP, PT, and TC are also broken. In fact, historically first was 
discovered that CP is broken and hence T should be broken as well.

The study of phenomenological manifestations of CPT violation has a long 
history. I will mention here only some selected contributions by
L.B. Okun~\cite{bogomol1-72}--\cite{lbo-02}. For recent works and review of 
the literature on CPT violation see refs~\cite{kostel1}--\cite{auriema}.

In what follows we consider baryogenesis, or more generally generation
of any cosmological charge asymmetry relaxing the assumption of CPT invariance.
For discussion of earlier works one may address review~\cite{ad-yaz}.
In what follows we reconsider and clarify the old results related to the
generation of charge asymmetry in thermal equilibrium due to mass difference 
between particles and antiparticles and discuss previously not considered case
of asymmetry when sacred principles of hermicity of the Hamiltonian
and thus unitarity of S-matrix or spin-statistics relation are broken.

According to CPT-invariance the masses of particles, $m$, and the corresponding 
antiparticles, $\bar m$, must be equal. If CPT is broken it is natural, though not
necessary, that this equality would be violated too and $m\neq \bar m$.
It is practically evident that in this case the number density of particles
and antiparticles may be unequal even in thermal equilibrium. Of course if 
baryonic charge or some other quantum number, $Y$, prescribed to particles is 
conserved, then the state with initially zero value of $Y$
would remain such in any evolution. We assume first for illustration
that the standard form of the
equilibrium distribution functions is not destroyed by CPT violation. This is
not necessarily true and the validity of this assumption is discussed below.
In equilibrium the particle distribution is described by the function:
\be
f(E,\mu) = \frac{1}{\exp[(E-\mu)/T] \pm 1},
\label{f}
\ee
where the signs ``$\pm$'' correspond to fermions and bosons respectively 
and $\mu$ is the chemical potential corresponding to quantum number $Y$.
For antiparticles in equilibrium with respect to annihilation
$\bar \mu = -\mu$. If the density of $Y$ is zero, then in 
CPT-invariant theory $\mu =0$. However, if $m\neq \bar m$, chemical potential
must be non-vanishing to ensure equality of particle densities, $n= \bar n$:
\be
\delta n \equiv n - \bar n = g_{df}\int \frac{d^3 p}{(2\pi)^3} \,\left[ f (E,\mu) - 
f(\bar E, \bar\mu) \right],
\label{delta-n-0}
\ee
where $E=\sqrt{p^2 + m^2}$, $\bar E =\sqrt{p^2 + \bar m^2}$,
and $g_{df}=g_s g_c g_g$ is the number of ``degrees of freedom'' of
the particle under scrutiny with $g_s$, $g_c$, and $g_g$ being the numbers of
the spins states, the number of colors, and the number of generations (families)
respectively.  
For example for three generations of quarks $g_{df} = 18$, corresponding
to 2 spin and 3 color states, for charged leptons $g_{df} =6$, and
for neutrinos $g_{df} = 3$, if the particle masses are smaller than temperature.

Evidently
if $\delta n =0$ but $m\neq \bar m$, chemical potential should be nonzero.
For sufficiently small mass difference, $\delta m = \bar m - m$,
such that $m\delta m/ET \ll 1$ we find: 
\be
\mu = \left({J_1}/{2 J_0}\right)\,  {m\delta m},
\label{mu}
\ee
where
\be 
J_0 = \int d^3 p f^2(E,0) \,e^{E/T}\,\,\,\, {\rm and}\,\,\,
J_1 = \int (d^3 p/ E) f^2 (E,0) \,e^{E/T}.
\label{J-0-J-1}
\ee
If, say, baryonic charge is not conserved and the processes of the
type $n + n \lrar {\rm mesons}$ or $(n - \bar n)$--oscillations 
are in equilibrium then the baryonic chemical
potential is forced to zero and there should be an excess of baryons over
antibaryons or vice versa in  thermal equilibrium,
\be
\delta n = g_{df}\,J_1 \,(m\delta m  /T)
\label{delta-n}
\ee

An interesting situation might be realized in the early universe
at the temperatures above the electroweak phase transition. As is generally
accepted, at such temperatures baryonic, $B$, and leptonic, $L$, numbers are
not conserved, while the difference $(B-L)$ is conserved, see e.g. 
reviews~\cite{ew-bs}. The processes with baryonic number violation include
colorless combination of all quarks and leptons from all three generations and 
lead in equilibrium to the following relation between chemical potentials:
\be
3(u+d) + (l +\nu) = 0,
\label{sum-mu}
\ee
where the particle symbol denotes the corresponding chemical potential,
$u$ and $d$ are respectively up and down quarks, $l$ is charged lepton,
$\nu$ is neutrino, and we
assumed that the chemical potentials do not depend upon the generation. 

Equilibrium with respect to the charged currents implies:
\be
W^+ = u - d = \nu - l.
\label{W+}
\ee
We do not distinguish here between chemical potentials of left and right-handed
fermions. Though it is a good approximation for quarks, due to their thermal
masses, it may be poorly valid for charged leptons, especially for electrons,
but nevertheless we neglect that for simplicity.

One more equation for determination of chemical potentials follows from the
condition of electro-neutrality:
\be
\frac{2}{3} \delta n_u - \frac{1}{3}\delta n_d - \delta n_l = 0,
\label{neutral}
\ee
where $\delta n$ is the difference of number densities of particles and
antiparticles. 

The last necessary equations follows from the fixation of the value of
the conserved density of $(B-L)$:
\be
\frac{1}{3}\left(\delta n_u + \delta n_d \right) - 
\delta n_l - \delta n_\nu = n_{B-L}\,.
\label{n-B-L}
\ee
Equations (\ref{sum-mu}) and (\ref{W+}) give:
\be
l = - (2u +d),\,\,\,\, \nu = -(u+2d),
\label{nu-l}
\ee
and hence from (\ref{neutral}) and (\ref{n-B-L}) follows:
\be
4\mu_u \left(J_{0u} + J_{0l}\right) - 2\mu_d \left(J_{0d} - J_{0l} \right)
&=& 2 m_u\delta m_u J_{1u} - m_d \delta m_d J_{1d} - m_l \delta m_l J_{1l},
\label{neutral-2} \nonumber \\
\mu_u\left( 2J_{0u} + 4 J_{0l} + J_{0\nu} \right)
+ 2\mu_d \left( J_{0d}  +  J_{0l} + J_{0\nu} \right) &=&
m_u\delta m_u J_{1u} + m_d \delta m_d J_{1d} - m_l \delta m_l J_{1l} \nonumber\\
&-& \frac{1}{2} m_\nu \delta m_\nu J_{1\nu} + {(2\pi)^3}\, T n_{B-L} 
\label{B-L1}
\ee
where $J_{i a}$ are given by equations (\ref{J-0-J-1}) with 
$E = \sqrt{p^2 + m_a^2}$ and
we have returned to the usual notations $u \rar \mu_d$, etc.

Solution of these equations is straightforward but tedious. We will present them 
only in the limit of high temperatures when
\be
J_0 = \frac{\pi^3 T^3}{6},\,\,\,\, 
J_1 = T^2 \ln 2
\label{J-hi-T}
\ee
independently on the particle type. Assuming equal masses and mass differences
for all quark generations we find 
after simple calculations for the baryon number density:
\be
n_B = - \frac{J_1}{(2\pi)^3 T}\left(\frac{9}{2} m_u \delta m_u +
\frac{15}{4} m_d\delta m_d \right)
\label{n-B}
\ee
and correspondingly the baryon asymmetry:
\be
\beta_T = \frac{n_B}{n_\gamma} = - 8.4\cdot 10^{-3} \left( 18 m_u\delta m_u + 
15 m_d \delta m_d \right) / T^2,
\label{beta}
\ee
where $n_\gamma = 0.24 T^3$ is the equilibrium number density of photons.
To take into account different masses of quarks from different families
$m\delta m$ should be changed into $\sum_a \delta m_a m_a/6$, where summation
is done over all quark families.  
We assumed above that there was no preexisting $(B-L)$ asymmetry and 
neglected lepton contributions. Below the 
electroweak temperature $T_{EW} \sim 100$ GeV baryonic number is practically
conserved and the asymmetry stays constant in the comoving volume up to the
entropy factor which diminishes $\beta$ by about an order of magnitude.
So to agree with the observed today value $\beta_T$ should be about $10^{-8}$. 
 
If we substitute the zero temperature values of the
quark masses into eq. (\ref{beta}) 
and take for an estimate an upper bound on $\delta m$ equal to 
the experimental limit on proton-antiproton mass difference,
$\delta m_p < 2\cdot 10^{-9}$ GeV~\cite{pdg},  
we see that the effect is by far too
small to explain the observed baryon asymmetry. Above the electroweak phase
transition Higgs condensate is absent and the vacuum masses of all fermions
are zero. However, there are significant temperature corrections to the masses,
$m(T) \approx g T$, where $g$ is the gauge (QCD) coupling constant. So the
high temperature quark masses are much larger than the lepton
masses. That's why we neglected above the lepton contributions into the baryon
asymmetry. 

To create the observed cosmological baryon asymmetry, $\beta_0 = 6\cdot 10^{-10}$,
we need $\delta m_q \sim (10^{-7} - 10^{-8} )T$ at $T\sim 100 $ GeV. It means that 
the quark mass difference should be about $10^{-5}-10^{-6} $ GeV, much larger
than the upper bound on the proton-antiproton mass difference. One would expect
$(m_p - m_{\bar p})$ to be of the same order of magnitude as $\delta m_q$.
An accidental cancellation is not excluded, but this looks quite unnatural. So we
have either to conclude that the
mass difference induced by CPT violation could not
create the observed baryon asymmetry or
to assume that the interaction which induces quark-antiquark mass difference
rises proportionally to temperature or characteristic energy scale
similarly to the usual quark masses. If this is the case then the expected
mass difference of proton and antiproton should be near $10^{-8}$ GeV not much
larger than the existing bound. An improvement of this limit by an order of
magnitude would exclude the mechanism of creation of the asymmetry
by the mass difference. Of course this statement is not rigorous because the
quark-antiquark mass difference may deviate much from that of 
proton-antiproton, but this looks rather unnatural, through not 
excluded in absence of the established theory. Another possibility
is that the quark-antiquark mass difference depends upon the quark flavor 
and may be much larger for $t$-quark than for the usual $u$ and $d$ quarks
which make nucleons.

All previous construction is heavily based on the assumption that the standard
form of the equilibrium distribution functions remain the same in CPT violating
theory. It is easy  to verify that distributions (\ref{f}) annihilate the
collision integral in T-invariant theory. Indeed the kinetic equation for 
the distributions of particles of type $j$ can be written as
\be 
\frac{df_j}{dt} = I_j^{(coll)},
\label{df-dt}
\ee
where the collision integral has the form
\be
I^{(coll)}_j = -\frac{1}{2E_j} \sum_f\int d\tau_j^{(in)} d\tau^{(fin)} 
(2\pi)^4\,\delta^{(4)} \left(\sum p_{in} -\sum p_{fin}\right) 
\left[ |A_{if}|^2 F_{if} - |A_{fi}|^2 F_{fi} \right],
\label{I-coll}
\ee
where the summation over $f$ is made over all possible final state particles,
$d\tau^{(fin)}$ is the phase space element of particles in the final state
and  $d\tau_j^{(in)}$ is the same for particles in the initial state with particle
$j$ excluded, 
$A_{if}$ and $A_{fi}$ are the amplitudes of transitions from initial to
final state and vice versa, and $F_{if}$ is the product of the distributions
of particles in the initial state and Fermi/Bose factors of those in the
final state:
\be
F_{if} = \Pi_i f_i \Pi_f (1\pm f_f)
\label{F-if}
\ee
The expression for $F_{fi}$ is obtained from $F_{if}$ by the interchange of the 
initial and final states.

In T-invariant theory $|A_{if}| = |A_{fi}|$ up to time reflections of kinematical 
variables which can be eliminated by a change of the integration variables. Hence the 
amplitude can be factored out from the square brackets in the r.h.s. of
eq.~(\ref{I-coll}). The remaining expression is proportional to 
$F_{if} - F_{fi}$, which vanishes for equilibrium functions (\ref{f}). It is
usually formulated as functions (\ref{f}) annihilate
collision integral because of the detailed
balance condition.

If we substitute equilibrium functions into collision integral in a theory 
which is not T-invariant, the integrand in $I_j^{(coll)}$ becomes proportional
to $[|A_{if}|^2 - |A_{fi}|^2]$, which is generally speaking non-zero.
So one may may worry if functions (\ref{f}) are the equilibrium ones
or not, i.e.
$I^{(coll)} \neq 0$, in T-violating theory, because the detailed balance condition
is violated. However, more general cyclic balance condition is 
fulfilled~\cite{ad-cycle} which leads to vanishing of the collision integral
on the usual equilibrium functions after summation over all possible reaction 
channels. It may be instructive to note that in the case that a single reaction 
channel is allowed, T-violation is not observable because it leads only to phase 
difference of T-conjugated amplitudes, $A_{fi} = \exp(i\theta) A_{if}$.
However an account of e.g. final state scattering destroys equality of absolute
values of T-conjugated amplitude and the effects of T-violation become
observable.

As is shown in ref.~\cite{ad-cycle}, the condition of vanishing of the collision 
integral after summation over all reaction channels follows either from
the unitarity of S-matrix or from conservation of probability plus CPT
invariance. So it is quite probable that if CPT invariance is broken, the 
equilibrium states would not be universal but would be different in different 
systems. However, such a strong conclusion about breaking of the usual 
equilibrium statistics
is not necessarily related to CPT breaking. Most probably if CPT is broken due
to breaking of the Lorenz invariance, equilibrium statistics does not change
but if e.g. CPT is broken due to non-hermicity of the Lagrangian, the 
equilibrium statistical distributions should be distorted too. Another possibility
is that if spin-statistics relation is broken, then almost surely the equilibrium
distributions would be different from the canonical ones because in this case
there is good chance that either locality or unitarity are broken
to say nothing of such ``minor'' things as breaking of CPT and Lorenz 
invariance~\cite{ad-venice}.

Baryogenesis in a scenario with spontaneously broken Lorenz invariance by
vacuum condensate of a tensor field was considered in paper~\cite{bertolami}.
The condensate breaks CPT invariance and leads to different energies of
particles and antiparticles and hence to a difference in their number densities.
The condensate acts in a similar way as the considered above mass difference
or a better analogy is that it induces different chemical potentials for particles and
antiparticles.

A connection between matter-antimatter asymmetry of the universe and possible
violation of CPT invariance was also discussed in ref.~\cite{gb-nm} in somewhat 
similar spirit. The authors studied CPT violating decoherence in neutrino 
oscillations induced by space-time foam. Resulting asymmetry between neutrinos
and antineutrinos could be transformed into baryon asymmetry by 
$(B+L)$-nonconserving electroweak processes.

It is difficult to make reliable estimates in non-existing frameworks of 
non-existing theory. At most one can hope for a reasonable guess. So
we will mimic violation of CPT which leads to distortion of the standard 
equilibrium distributions assuming that the detailed balance is broken in 
the collision integral with a 
single channel allowed (which, as we have mentioned above,
contradicts unitarity in a normal theory). So we assume that
\be
|A_{fi}|^2 = |A_{if}|^2 \left( 1 + \Delta_{if} \right).
\label{Delta-if}
\ee
Let us consider as an example the process $a_1+a_2 \lrar a_3 + a_4$, where
$a_j$, for definiteness, though not obligatory, are fermions.
We assume that the equilibrium distributions are only slightly modified
and write $f_j^{(eq)} = f_j(1+\delta_j)$, where $f_j$ is the standard
equilibrium function given by eq. (\ref{f}).
The distribution of, say, particle $a_1$  evolves according to:
\be
f_1 \dot \delta_1 =\frac{1} {2E_1} \int d\tau_2 d\tau_{34} (2\pi)^4
\delta^{(4)} \left( p_1+p_2 -p_3-p_4\right) |A_{12}|^2
f_1 f_2 (1-f_3)(1-f_4) \nonumber \\
\left[ \Delta  +  \delta_3 \frac{2-f_3}{1-f_3} + \delta_4 \frac{2-f_4}{1-f_4}-
\delta_1 \frac{2-f_1}{1-f_1} -\delta_2 \frac{2-f_2}{1-f_2} 
\right],
\label{dot-delta-f}
\ee
where we have omitted the indices difference of the 
amplitudes, $\Delta$, induced by CPT-breaking,
and assumed that the system is stationary so the temperature is
constant and the usual equilibrium distributions are time independent.
It is straightforward to generalize the equation to the case of time varying temperature
and chemical potentials.
We have neglected effects of the particle-antiparticle mass difference. Since they 
enter linearly for small $\delta m$, it is easy to include them using the derived
above equations. Note that the equation for $\delta_3$ differs from 
eq. (\ref{dot-delta-f}) by the sign of the expression in square brackets.
It is evident that for elastic scattering amplitude $\Delta = 0$ and if only
elastic scattering is essential then all $\delta_j =0$.

Equilibrium is defined as the state for which $\dot\delta_j = 0$. Now the factor in
square bracket cannot be zero because each function $f_j$ and $\delta_j$
depends only on $E_j$. Even in the limit of Boltzmann statistics, when $f_j \ll 1$
this may not be realized because of possible non-trivial dependence of $\Delta$
on the energies of the participating particles. Anyhow it is evident that equilibrium
distributions cannot be universal but depend upon the concrete participating 
reactions.

Another interesting feature is that the equilibrium is not realized locally 
for an arbitrary value of the particle energy but only on the average integrated over 
phase space of the reactions. There are four
integral relation for $\delta_j (E_j)$, $j=1,2,3,4$. In the oversimplified case
when $|A_{12}|^2$ and $\Delta$ are constant and $f_j \ll 1$ there is a trivial, 
though non-realistic, solution $\delta_1 =\delta_2 = -\delta_3 = -\delta_4 = \Delta/4$.
It is natural to expect that $\Delta$ is a function of kinematic variables,
$\Delta =\Delta (s,t)$, where $s=(p_1+p_2)^2$ and
$t= (p_1-p_2)^2$. In this case local equilibrium is impossible.

Most probably the equilibrium point, where $\dot\delta_j =0$, is a stable one and
the solution tends to a time-independent limit. If this is the case then CPT and
unitarity violations do not lead to effects which accumulate with time.

Returning to the cosmological baryon asymmetry, we should expect that the corrections
to the equilibrium distribution functions for particles and antiparticles are different
and so are their number densities. The asymmetry in this case is not directly
related to the mass difference of particles and antiparticles and so is not 
bounded by the experimental limit on $\delta m$.


\end{document}